\documentclass[aps,amsmath,amssymb,preprintnumbers,groupedaddress,twocolumn]{revtex4}
\usepackage{graphicx,psfrag,hyperref}

\def\hepth#1{\href{http://xxx.arxiv.org/abs/hep-th/#1}{{arXiv:hep-th/#1}}}

\def\arxiv#1#2{\href{http://xxx.arxiv.org/abs/#1}{{arXiv:#1 [#2]}}}

\newcommand{\beq}{\begin{equation}}
\newcommand{\eq}{\end{equation}}
\newcommand{\req}[1]{(\ref{#1})}
\newcommand{\ie}{{\it i.e.}}

\renewcommand{\t}{\tilde}

\newcommand{\Fcal}{\mathcal F}

\newcommand{\Ncal}{\mathcal N}

\newcommand{\Ccal}{\mathcal C}
\newcommand{\Bcal}{\mathcal B}

\newcommand{\m}{m}

\newcommand{\ep}{\epsilon}

\newcommand{\corr}[1]{\left<#1\right>}

\newcommand{\cf}{{\it cf.}}

\begin{document}

\preprint{UCB-PTH-12/15}
\title{Penner Type Ensemble for Gauge Theories Revisited}

\author{Daniel Krefl$^{a}$}
\affiliation{\it ${}^{a}$ Center for Theoretical Physics, University of California, Berkeley, USA
}

\begin{abstract}
The Penner type $\beta$-ensemble for $\Omega$-deformed $\Ncal=2$ $SU(2)$ gauge theory with two massless flavors arising as a limiting case from the AGT conjecture is considered. The partition function can be calculated perturbatively in a saddle-point approximation. A large $N$ limit reproduces the gauge theory partition function expanded in a strong coupling regime, for any $\beta$ and beyond tree-level, confirming previous results obtained via special geometry and the holomorphic anomaly equation. The leading terms and gap of the gauge theory free energy at the monopole/dyon point follow as a corollary.
\end{abstract} 
\maketitle
\noindent

\section{Introduction}
In recent years there has been continuous progress in understanding and extending the AGT conjecture, formulated in \cite{AGT09}, which relates two dimensional conformal field theory and four dimensional gauge theory. 

One of the central statements of the original conjecture is the equivalence between conformal blocks, $\Bcal$, and the instanton part of $\Omega$-deformed $\Ncal=2$ gauge theory partition functions, $Z^{inst}$, \ie,
\beq\label{BvsZ}
\Bcal_\gamma(\alpha_i)\sim Z^{inst}(a_i,m_i,q_i;\beta)\,,
\eq
which invokes a non-trivial mapping between the gauge theory parameters, namely the Coloumb moduli $a_i$, masses $m_i$ of matter fields, UV gauge couplings $q_i$ and the $\Omega$-deformation parameter $\beta:=-\ep_2/\ep_1$, and the conformal dimensions $\Delta_i=\alpha_i(\alpha_0-\alpha_i)$ and positions of the operator insertions on the CFT side. (Note that there should be as well a $g_s$ dependence in \req{BvsZ}, since $\ep_1=\sqrt{\beta}g_s$ and $\ep_2=-\frac{1}{\sqrt{\beta}}g_s$ is the natural parameterization for the equivariant parameters $\ep_i$. However, for historical reasons $g_s$ will only be introduced later.) Here, $\alpha_0$ is given in terms of the central charge $c$ via
$$
c=1-24\alpha_0^2\,,
$$
and $\gamma$ denotes the set of intermediate states. 

In this note we are in particular interested in the $\Omega$-deformed $\Ncal=2$ asymptotically free $SU(2)$ gauge theories with $N_f\leq 3$, which can be obtained as a limiting case  of the theory with $N_f=4$. The latter is identified via the relation \req{BvsZ} with the sphere 4-point conformal block with a $s$-channel intermediate state under the identification of parameters \cite{AGT09}
\beq\label{CFTtoGauge}
\begin{split}
&\gamma=\alpha_0+a\,,\,\,\,\,\,\alpha_0=\frac{1}{2}\left(\sqrt{\beta}-\frac{1}{\sqrt{\beta}}\right)\,,\\
&\alpha_1=\alpha_0+\frac{1}{2}\left(m_1-m_2\right)\,,\,\,\,\,\,\alpha_2=\frac{1}{2}\left(m_1+m_2\right)\,,\\
&\alpha_3=\frac{1}{2}\left(m_3-m_4\right)\,,\,\,\,\,\,\alpha_4=\alpha_0+\frac{1}{2}\left(m_3+m_4\right)\,.
\end{split}
\eq
(The $m_i$ are the masses of the hypermultiplets.) Decoupling of flavors extends \req{BvsZ} to a correspondence between so-called irregular conformal blocks and the $N_f\leq 3$ theories \cite{G09,MMM09}.

A remark is in order. The identification \req{BvsZ} for the asymptotically free theories has been originally inferred for expansion at weak IR coupling, \ie, with large $a$, respectively $\gamma$. However, one should keep in mind that one should see in \req{BvsZ} more than a mere statement for weakly coupled gauge theory. In general, and in particular for the theories of our interest, $SU(2)$ with $N_f\leq 3$, the gauge theory partition function $Z^{inst}$ is a non-trivial function over the Coloumb parameter moduli space. This is reflected in the non-uniqueness of the conformal blocks $\Bcal$. There is an underlying monodromy problem, that is, the blocks have (qualitatively said) non-trivial monodromies
$$
\Bcal_\gamma\rightarrow M_{\gamma l}\,\Bcal_l\,.
$$ 
Hence, a fully fledged AGT conjecture for the asymptotically free theories includes a mapping between the underlying monodromy problems, as the relation $\gamma \sim a$ suggests. In other words, for each $Z^{inst}$ expanded in a specific region in Coloumb parameter moduli space, there should be a corresponding representation of the conformal block. 

Therefore, for fully fledged confirmations and/or proofs of the relation \req{BvsZ} one has to go beyond the regime of IR weakly coupled gauge theory, and in particular discuss the mapping of the underlying monodromy problems.  This appears to have been discussed only superficially so far, if at all (however, certain aspects can be inferred from \cite{CDV10}).

Therefore, one might see the proofs of some specific instances of the AGT conjecture available today as sort of incomplete, as they appear to be valid only at one specific point in moduli space (\cf, \cite{HJS10}). In particular, most of the explicit confirmations of the conjecture are based on matching CFT calculations for $\Bcal$ with the instanton counting formulae of \cite{N02}. 

One of the reasons for the lack of confirmations of the AGT conjecture at other points in moduli space might be that so far there are no such neat expressions for $Z^{inst}$ expanded at strong coupling available as the formulaes of \cite{N02}. However, techniques to obtain $Z^{inst}$ for the $N_f\leq 3$ theories at other points in moduli space are available, though technically more involved. They make use of special geometry and the (extended) holomorphic anomaly equation, as put forward in \cite{KW10a,KW10b}. This is usually referred to as B-model approach, as the underlying techniques have been originally developed to calculate B-model topological string amplitudes. 

One of the purposes of this little note is to show at hand of an explicit example that the relation $\req{BvsZ}$ indeed continues to hold away from weak IR coupling for the asymptotically free theories. Fortunately, many of the needed ingredients are already scattered in the literature.

The example we will focus on in this note is $\Omega$-deformed $\Ncal=2$ $SU(2)$ gauge theory with two massless flavors. The reason being that both, the B-model calculations, as well as the calculation of the corresponding blocks are rather simple. Since we will utilize the Dotsenkov-Fateev integral representation of the sphere 4-point conformal blocks, the correspondence \req{BvsZ} translates to the proposed equivalences of \cite{DV09} between $\Omega$-deformed $SU(2)$ partition functions with $N_f\leq 4$ and $\beta$-ensembles of Penner-type.

Hence, the confirmation of \req{BvsZ} we perform below is a non-trivial higher genus check of the $\beta$-ensemble as the incorporation of the $\Omega$-deformation. To our knowledge, this is the first check of a (non-gaussian) $\beta$-ensemble beyond genus one. It is expected that the results extend to the massive $N_f=2$ and $N_f=3$ cases. (In fact, it seems possible to write down a proof of the AGT conjecture for the asymptotically free theories for any $\beta$ and for any point in Coloumb moduli space using the known techniques featured in this note, though certain details still have to be worked out.) 

Finally, let us mention one of the reasons why a topological string theorist might be interested in the validity of the relation \req{BvsZ} beyond weak coupling. 
CFT calculations of conformal blocks can be rather simple (see in particular \cite{G09}), and so, hopefully, their analytic continuation. Some indications in this direction can be found below. Since there are extensions of the AGT correspondence to 5D \cite{AY09a,AY10} (from the ensemble point of view, this will involve a $q$-deformation of the measure), one might hope that one can similarly calculate the 5D partition functions away from weak coupling via CFT, \ie, reach other points in moduli space of the corresponding refined topological string. This will yield independent confirmations of refined B-model calculations, for which no other way of calculation is known so far.

\section{From Blocks to Ensembles}
In this section we will present a derivation of the Penner type $\beta$-ensemble corresponding according to the AGT conjecture to $\Omega$-deformed $SU(2)$ gauge theory with $N_f=4$. The derivation is in spirit the one of \cite{DV09,MMS09,MMS10}, though we will present things in a slightly different (and more condensed) form.

Starting point is the representation of the conformal block of the sphere 4-point correlation function in terms of the free field correlator with screening charge insertions (also known as Dotsenkov-Fateev integral) \cite{DF84a,DF84b}
\beq\label{DFint}
\begin{split}
&\Bcal_\gamma(\alpha_i;q)\sim\oint_{[\Ccal^+_\gamma]}[d\lambda] \oint_{[\Ccal^-_\gamma]}[d\t \lambda]\\
& \times\corr{V_{\alpha_1}(\infty)V_{\alpha_2}(1)V_{\alpha_3}(q)V_{\alpha_4}(0)\prod_{i=1}^{S_+}J_+(\lambda_i)\prod_{j=1}^{S_-}J_-(\t \lambda_j)}\,,
\end{split}
\eq
where $[\Ccal^\pm_\gamma]$ denotes the set of integration contours, $V_{\alpha_i}(z_i)$ is an usual vertex operator inserted at position $z_i$ represented as an exponent of a free field and $J_\pm:=V_{\alpha_\pm}$.
One should note that different choices of contours $[\Ccal^\pm_\gamma]$ lead to different solutions for $\Bcal$, reflecting the non-uniqueness of the blocks (\ie, the monodromy problem), \cf, \cite{DF84b}. For instance, a choice of contours corresponding to the weakly coupled gauge theory has been proposed in \cite{MMS10,IO10}. Here, we will be naturally led to a different choice of contours (which should actually be the one of \cite{DV09}). Note that we neglected in \req{DFint} an overall ($q$-dependent) factor, which is not of relevance for our specific example later on.

The free field correlator \req{DFint} is supplemented by the neutrality condition 
\beq\label{ffcondition}
\sum_{i=1}^4\alpha_i+S_+\alpha_+ +  S_-\alpha_- =2\alpha_0\,,
\eq
with
$$
\alpha_\pm=\alpha_0\pm \sqrt{\alpha_0^2+1}\,.
$$ 
In particular, for $\alpha_0$ given in \req{CFTtoGauge}, one has 
\beq\label{alpha0pm}
\alpha_\pm=
\left\{
\begin{matrix}
\sqrt{\beta}&\\
-\frac{1}{\sqrt{\beta}}
\end{matrix}
\right.\,.
\eq

Due to the mapping of parameters \req{CFTtoGauge}, we want to keep $\alpha_i$ unconstrainted (that is, we want to allow arbitrary masses). Hence, the relation \req{ffcondition} imposes a condition of the number of screening charge insertions. Clearly, one set of screening operators is sufficient to fulfill \req{ffcondition}. Let us pick $J_+$. We set $S_-=0$ and define $N:=S_+$ such that we obtain from \req{DFint} the eigenvalue ensemble
\beq\label{Pensemble}
\begin{split}
&\Bcal_\gamma(\alpha_i;q)\sim\\
&\oint_{[\Ccal_\gamma^+]}[d\lambda] \Delta(\lambda)^{2\alpha^2_+}\prod_{i=1}^N(\lambda_i)^{\alpha_+\alpha_4}(\lambda_i-1)^{\alpha_+\alpha_2}(\lambda_i-q)^{\alpha_+\alpha_3}\,,
\end{split}
\eq
with $\Delta(\lambda):=\prod_{i<j}^N(\lambda_i-\lambda_j)$ the usual Vandermonde, and condition on the number $N$ of eigenvalues 
\beq\label{Ncond}
N=-\frac{1}{\alpha_+}\left(\sum_{i=1}^4\alpha_i-2\alpha_0\right)=-\frac{1}{\sqrt{\beta}}\left(m_1+m_3\right)\,.
\eq
For brevity of notation, we define $Z:=\Bcal_\gamma(\alpha_i;q)$ for later usage. 


The single product in \req{Pensemble} can be written as a logarithmic potential, \ie, as $e^{\sum_{i=1}^N W(\lambda_i)}$ with
$$
W(\lambda)=\alpha_+\left(\alpha_1\log(\lambda)+\alpha_2\log(\lambda-1)+\alpha_3\log(\lambda-q)\right)\,.
$$
Therefore the ensemble \req{Pensemble} is usually referred to as of Penner type. We still have to bring in an additional overall $g_s^{-1}$ in the potential ($g_s$ is needed to match the ensemble with a small $g_s$ expansion of the $\Omega$-deformed gauge theory). Since we do not want a $g_s$ dependence in the measure (and also not in the central charge), we do not redefine $\alpha_0$, but instead the masses $m_i\rightarrow m_i/g_s$. In particular, under this redefinition the $\alpha_0$ terms occuring in the $\alpha_i$ given in \req{CFTtoGauge} turn into quantum shifts of the mass parameters in the ensemble. Hence, we arrive at the $\beta$-ensemble proposed in \cite{DV09}.

In the limit $g_s\ll 1$ we can perform a saddle-point expansion of the ensemble as the potential possesses two distinct critical points, say $\mu_\pm$. Hence, in the small $g_s$ limit a two cut structure emerges, which fixes the set of integration contours $[\Ccal^+]$, as the eigenvalues localize around the critical points. Denoting the number of eigenvalues being located near $\mu_\pm$ as $N_\pm$, with $N=N_++N_-$, we can take in addition a large $N_\pm$ limit, \ie,
$$
N_\pm\rightarrow\infty\,, 
$$
In this limit we keep
\beq\label{adef}
a_\pm:= \sqrt{\beta} g_s N_\pm\,,
\eq
fixed. The condition \req{Ncond} then translates to a relation between $a_+$ and $a_-$, leaving one parameter, which we denote as $a$, unconstrainted. As the notation suggests, this parameter will be identified with the Coloumb modulus of $\Ncal=2$ $SU(2)$ gauge theory. However, since $a$ is finite, we expect that generally we are sitting in a strongly coupled point in the gauge theory moduli space.  In the following we will confirm this at hand of the (calculationally) simplest example, namely the case with massless $N_f=2$ obtainable by decoupling two of the flavors and sending the remaining masses to zero.  Note that the massless $N_f=2$ case is nicely behaved in the sense that one can perform the decoupling and massless limit directly on the level of the $\beta$-ensemble, as the two cut structure survives the massless limit (the $N_f=3$ case is more subtile). 

\section{Ensemble for massless $N_f=2$}
The decoupling limit of flavors down to $N_f=2$, following \cite{SW94b}, directly applied on the level of the ensemble is straight-forward and has been already discussed to some extend in the literature \cite{EM09a,EM09b}. In detail, sending $m_4\rightarrow \infty$ keeping $m_4q=\Lambda_3$ fixed and subsequently $m_2\rightarrow \infty$ with $\m_2\Lambda_3=\Lambda_2^2$ fixed (this requires also a rescaling $\lambda\rightarrow \frac{\Lambda_3}{\Lambda_2}\lambda$), yields the rather simple potential (with remaining matter massless) 
\beq\label{WmasslessNf2}
W(\lambda)=-\frac{\Lambda_2}{2}\left(\lambda+\frac{1}{\lambda}\right)\,,
\eq
where $\Lambda_2$ denotes the dynamical scale (this also involved taking the UV coupling $q$ small). Furthermore, the condition on the number of eigenvalues \req{Ncond} is reduced to
\beq\label{Nf2Ncond}
N=0\,.
\eq
(We picked here a specific quantum shift (gauge) of mass parameters which simplifies the calculations. One should note that the precise choice of gauge is not of relevance for the statement \req{BvsZ}.)
The reduced condition \req{Nf2Ncond} looks puzzling at first sight, since it seems to dictate that no eigenvalues are allowed to be present! However, as discussed above, a double scaling limits circumvents this. (In fact, a saddle-point expansion ($g_s\ll 1)$ of the ensemble alone is sufficient if one allows for a negative number of eigenvalues, which one should see as eigenvalue holes.) Clearly, the potential \req{WmasslessNf2} possesses two critical points $\mu_\pm=\pm 1$, and we denote as before the number of eigenvalues being located around $\mu_\pm$ as $N_\pm$. Then, the condition \req{Nf2Ncond} translates to $N_-=-N_+$.  In particular, we have that
$$
a:=a_+=-a_-\,,
$$
with $a_\pm$ as defined in \req{adef}.

As usual, in the saddle-point approximation the ensemble partition function splits into two parts 
\beq\label{Zsplit}
Z=Z_{np}\,Z_{pert}\,.
\eq
The perturbative part, $Z_{pert}$, is expressible as a sum of normalized gaussian correlators, efficiently calculable in a small $g_s$ expansion, following for instance \cite{KMT02,MS10}. This yields $$
\log Z_{pert}\sim \sum_{g=0}^\infty\Fcal^{(g)}_{pert}(a;\beta)\,\left(\frac{g_s}{\Lambda_2}\right)^{2g-2}\,,
$$
with 
\beq\label{Fpert}
\Fcal^{(g)}_{pert}(a;\beta):=\sum_{n=1}^\infty c^{(g)}_n(\beta)\, a^{n}\,, 
\eq 
and $c^{(g)}_n(\beta)$ $\beta$-dependent coefficients, some of them given for the reader's convenience in tables \ref{cgtableTree} and \ref{cgtable}.\begin{table}
\begin{center}
\begin{tabular}{c|c|c|c|c|c|c|c|}
n/g&1&2&3&4&5&6&7\\
\hline
0&$0$&$0$&$-\frac{1}{2}$&$\frac{5}{16}$&$-\frac{11}{32}$&$\frac{63}{128}$&$-\frac{527}{640}$\\
\end{tabular}
\caption{Coefficients $c_n^{(g)}$ of order $g_s^{-2}$ for some low $n$.}
\label{cgtableTree}
\end{center}
\end{table}
\begin{table*}
\begin{center}
\begin{tabular}{c|c|c|c|}
n/g&1&2&3\\
\hline 
1&$\frac{3-7\beta+3\beta^2}{4\beta}$&$-\frac{17-41\beta+17\beta^2}{16\beta}$&$\frac{205-503\beta+205\beta^2}{96\beta}$\\
2&$\frac{21-88\beta+131\beta^2-88\beta^3+21\beta^4}{64\beta^3}$&$-\frac{297-1198\beta+1736\beta^2-1198\beta^3+297\beta^4}{128\beta^3}$&$\frac{7574 - 29660 \beta + 42025 \beta^2 - 29660 \beta^3 +7574 \beta^4}{640\beta^3}$\\
\end{tabular}
\caption{Expansion coefficients $c_n^{(g)}$ of order $g_s^0$ and $g_s^2$ for some low $n$.}
\label{cgtable}
\end{center}
\end{table*}

Note that necessarily $\Fcal^{(g)}_{pert}$ is a polynomial in $a$, since the gaussian correlators are polynomials in $N$.

On the other hand, $Z_{np}$ is determined by the normalization of the correlators, which is simply given by the gaussian, \ie,
$$
Z_{np}=Z_g(N_+)\times Z_g(N_-)\,,
$$
with 
$$
Z_g(N)=\int[d\lambda]\Delta(\lambda)^{2\beta} \, e^{-\frac{1}{2}\sum_{i=0}^{N}\lambda_i^2}=\prod_{i=1}^N\frac{\Gamma(1+\beta i)}{\Gamma(1+\beta)}\,.
$$
It is well known that the gaussian free energy $\log Z_g(N)$ possesses the large $N$ expansion, 
$$
\log Z_{g}(N)\sim \dots+ \sum_{n\geq0}^\infty \Phi^{(n)}(\beta)\, \frac{1}{N^n}\,,
$$ 
with expansion coefficients $\Phi^{(n)}$ given by (\cf, \cite{KW10a}) 
\beq
\begin{split}
&\Phi^{(0)}(\beta)=\frac{1}{12}\left(\beta+\frac{1}{\beta}\right)-\frac{1}{4}\,,\\
&\Phi^{(n>0)}(\beta)=(n-1)!\sum_{k=0}^{n+2}\frac{(-1)^kB_kB_{n+2-k}}{k!(n+2-k)!}\beta^{k-n/2-1}\,,
\end{split}
\eq
where $B_n$ denotes the $n$-th Bernoulli number. Hence, using \req{adef} we infer that
\beq\label{Fnp}
\log Z_{np}(a;g_s,\beta)=\dots+2\sum_{n>0}^\infty\Phi^{(2n)}(\beta)\, \left(\frac{g_s}{a}\right)^{2n}\,.
\eq
Combining \req{Fpert} with \req{Fnp} we deduce that the free energy $\log Z$ for the potential \req{WmasslessNf2} possesses the characteristic gap structure observed for $SU(2)$ with $N_f\leq 3$ massless flavors expanded near a massless monopole/dyon point in moduli space \cite{KW10a,KW10b}, with some distinguished leading singular terms $\Phi^{(n)}$. In particular, the $\Phi^{(n)}$ are related to the expansion coefficients of the $c=1$ string free energy (at radius $R=\beta$), yielding the universal leading singular terms for the gauge theory expanded at a monopole/dyon point, via a shift of $N$ \cite{KW10b}.

Therefore, we expect that the large $N_\pm$ limit of $Z$ corresponds to the $\Omega$-deformed $SU(2)$ gauge theory partition function with massless $N_f=2$ expanded near a monopole/dyon point, but with some additional shift of parameters.

Note that the (gap) structure \req{Zsplit} is independent from the specific model under consideration, \ie, it will continue to hold for massive $N_f=3$ and $N_f=2$, and hence also for the massless limits all the way down to pure $SU(2)$.  (This is in fact a general property of $\beta$-ensembles in the saddle-point approximation with multiple cuts, as for instance observed already for ordinary matrix models ($\beta=1$) in \cite{AKMV02}).

\section{B-model verification}
The free energy of $\Omega$-deformed $SU(2)$ gauge theory with $N_f\leq 3$ (massive or massless), can be obtained for any quantum shift of masses and at any point in moduli space via invoking the (extended) holomorphic anomaly equation of \cite{BCOV93a,BCOV93b,W07} (the B-model approach). We will not recall here all the details, but rather refer to the works \cite{KW10a,KW10b,HKK11} and references therein instead. 

According to \req{BvsZ}, and the previous two sections, the coefficients of the expansion \req{Fpert} at order $g_s^{-2}$, given in table \ref{cgtableTree}, should correspond to the expansion of the prepotential of $\Ncal=2$ $SU(2)$ gauge theory with two massless flavors near a monopole or dyon point. This can be easily verified via the Seiberg-Witten solution of $\Ncal=2$ gauge theory \cite{SW94a,SW94b}. In detail, the period $a(u)$ and its dual $a_D(u)$ are obtained at a specific point in quantum moduli space (parameterized by $u$) via solving a corresponding Picard-Fuchs equation. The prepotential follows from the well known relation
$$
a_D=\frac{\partial\Fcal^{(0)}}{\partial a}\,.
$$
We observe that the instanton terms of order $a^{>2}$ of the prepotential expanded at a monopole/dyon point can be matched with the coefficients obtained from the ensemble listed in table \ref{cgtableTree} (here, and in the following, a matching of used conventions might require a rescaling of $\Lambda_2$). 

Higher order terms in $g_s$ correspond to gravitational corrections to the gauge theory and can be calculated using the holomorphic anomaly equation.
One of the essential points in that approach  being  that one has to supplement by hand holomorphic functions, the so-called holomorphic ambiguity. This is usually done by assuming specific leading terms of the free energies expanded near the dyon/monopole point, following the ideas of \cite{HK06} in the case of (non $\Omega$-deformed) $SU(2)$. 

Here, things are however a little bit more tricky since we do not know the precise quantum shift (gauge) of parameters in the ensemble \req{Pensemble}, which becomes important beyond tree-level (in particular shifts of $N_\pm$). Therefore, we proceed in the inverse way of \cite{KW10a}. That is, we fix at the dyon/monopole point the anomaly to match the ensemble results given in table \ref{cgtable} and analytically continue back to the weak coupling regime to compare with the instanton calculus of \cite{N02}. 

The 1-loop sector is rather simple. We infer that
\beq\label{F1Bmodel}
\Fcal^{(1)}=-\frac{1}{2}\log\left(\partial_u a(u)\right) -2\,\Phi^{(0)}(\beta) \log(64u^2-\Lambda_2^4)\,,
\eq
reproduces the coefficients $c^{(1)}_n$ (with $u(a)$ at the monopole/dyon point). Since we know $a(u)$ at any point in moduli space, \req{F1Bmodel} can be easily expanded as well at weak coupling ($a\rightarrow\infty$). Comparison with the instanton calculus of \cite{N02} then shows that the 1-loop sector is in accord with the massless limit of Nekrasov's original $N_f=2$ partition function under the additional quantum shift (this is not the shift used in \cite{KW10b}, but precisely the shift we used to obtain the simplified potential \req{WmasslessNf2} with condition \req{Nf2Ncond})
\beq\label{shift}
\begin{split}
m_1&\rightarrow m_1+\frac{1}{2}\left(\sqrt{\beta}-\frac{1}{\sqrt{\beta}}\right)g_s\,,\\
m_2&\rightarrow m_2-\frac{1}{2}\left(\sqrt{\beta}-\frac{1}{\sqrt{\beta}}\right)g_s\,,
\end{split}
\eq
before sending $m_i\rightarrow 0$. 

Higher $\Fcal^{(g)}$ can be calculated recursively via the holomorphic anomaly equation of \cite{BCOV93b} (respectively, of \cite{W07}, depending on choice of gauge of mass parameters \cite{KW10b}). We do not give any more details here, but just state that the holomorphic anomaly of $\Fcal^{(g>1)}$ can be fixed such that at the dyon/monopole point we reproduce the $\Fcal^{(g)}$ from the ensemble. Analytic continuation to the weakly coupled regime shows again accordance with the instanton calculus under the mass shift \req{shift}. We dared to check up to order $g_s^4$ (only a small subset of the obtained coefficients $c^{(g)}_n$ is shown in table \ref{cgtable} because the $\beta$-dependent expressions quickly become rather lengthy).

This confirms that \req{BvsZ} is valid away from weak coupling for the example under consideration. In particular, the validity of the Penner-type $\beta$-ensemble as a dual to $\Omega$-deformed gauge theory is confirmed, for any $\beta$ and beyond tree-level. The leading terms at the monopole/dyon point in moduli space (given by the $c=1$ string partition function at $R=\beta$), and the gap condition found for $\Omega$-deformed gauge theory in \cite{KW10a,KW10b} follow as a corollary (mainly due to \req{Zsplit} and polynomiality of gaussian correlators).

\acknowledgments
The work of D.K. has been supported by a Simons fellowship, and by the Berkeley Center for Theoretical Physics.


\end{document}